\documentclass[a4paper,11pt]{article}
\pdfoutput=1 

\usepackage{jcappub} 

\usepackage[T1]{fontenc} 
\usepackage{caption}
\usepackage{subcaption}
\usepackage{braket}
\usepackage{amsfonts}
\usepackage{booktabs}

\title{\boldmath Uncertainties in measuring the dark matter signal from Milky Way satellites using Cherenkov telescopes}

\author[a,1]{Maria Kherlakian,\note{Corresponding author.}}
\author[a]{Aion Viana,}
\author[a]{Vitor de Souza}

\affiliation[a]{Instituto de F\'isica de S\~ao Carlos, Universidade de S\~ao Paulo, Av. Trabalhador S\~ao-Carlense 400, S\~ao Carlos, Brasil.}

\emailAdd{maria.kherlakian@desy.de}
\emailAdd{aion.viana@ifsc.usp.br}
\emailAdd{vitor@ifsc.usp.br}

\abstract{In this work, we present a modelling of the galactic sub-clumps based on statistical estimations of the full Milky Way satellite population. We introduce 10 substructure modellings (SM$_{i}$, $i \in \{1, …, 10\}$) with the following varying parameters: a) subhalos inner profile, b) spatial distribution of subhalos, c) mass distribution of subhalos, d) total number of subhalos and e) concentration parameter. The sensitivity curves of CTA for sources in each model are calculated for the $\tau^{+}\tau^{-}$ and $b\bar{b}$ decay channels. With both detection of a signal (5$\sigma$) with the CTA and no signal observation, no model was effective in accessing the thermal values of $\braket{\sigma v}$. We analyse the systematic effect introduced by the substructures models.}

\begin{document}
\hspace{-1cm} https://doi.org/10.1088/1475-7516/2023/03/025   \hspace{3cm}  JCAP03(2023)025
\maketitle
\flushbottom
\section{Introduction}
\label{sec:intro}
The nature of dark matter (DM) is still one of the biggest mysteries of modern science. Unveiling its complexion through indirect observations of astrophysical gamma-rays requires state-of-the-art observatories, such as the Cherenkov Telescope Array (CTA)~\cite{cta2017}. Within the Milky Way (MW), DM is distributed as one main halo and a large population of orbiting subhalos. The largest subhalos are believed to belong to the dwarf spheroidal galaxies (dSphs) orbiting the MW. Studies based on dynamical mass models of these galaxies fitted to kinematical data of their stars predict they are among the most dark-matter dominated objects. Because of their low astrophysical background in high energies and proximity, they are also promising targets for DM indirect searches. The present status of known MW satellites, however, is partially incomplete. The number of dSphs as of August 2020 stands at 56: 17 discovered by the Sloan Digital Sky Survey SR9 (SDSS)~\cite{sdss}, 17 by the Dark Energy Survey (DES)~\cite{des1, des2}, 11 classical dSphs and 11 discovered by other surveys. Because of the limited sky coverage and sensitivity of sky-survey experiments and results of N-body simulations~\cite{Aquarius, elvis}, it is believed that this number is much larger.

Besides those, a larger number of small DM concentrations, known as dark subhalos, might orbit the MW~\cite{Zavala:2019gpq}. Dark subhalos are almost invisible in the electromagnetic spectrum because of their low baryonic content. Gravitational evidences of the existence of dozens of dark subhalos have been detected~\cite{Zavala:2019gpq} and N-body simulations~\cite{Aquarius, elvis} predict a number thousands of times higher. They should also contribute to the total dark matter signal to be measured by CTA. Strategies for detecting dark subhalos are present in~\cite{Khlopov} and were studied in particular with CTA in references~\cite{Coronado-Blazquez:2021avx, dmhutten}. Our work introduces the effects of the expected total Milky Way massive satellite population on the prospects of dark matter detection with the CTA. Within the N-body cosmological simulation framework Via Lactea II, the detectability can be similar to the one expected from dwarf galaxies.

In this paper, we focus on the unknown parameters and statistical uncertainties of the Milky Way subhalo population models. N-nody simulations have several parameters which need to be set to data and, after optimization, they represent the most probable outcome of a wide range of possibilities. We define 10 subhalo models varying: a) subhalos inner profile, b) spatial distribution of subhalos, c) mass distribution of subhalos, d) total number of subhalos and e) concentration parameter. For each model, a detailed simulation of the gamma-ray signal and how it would be detected by CTA is presented.

In section~\ref{sec:models}, we define the models used in the paper. In section~\ref{sec:sim}, we explain how the simulation of the sources and detector is taken into account. We simulate the detection of the sources by the CTA observatory. In section~\ref{sec:results}, the results are shown, illustrating detection and upper limits. Section~\ref{sec:conclusion} concludes the work.

\section{Definition of dark matter subhalos models}
\label{sec:models}

Given a main dark matter halo, its substructure population can be fully characterised by the differential number of subhalos in elements of volume, mass and concentration parameter:

\begin{equation}
\frac{d^{3}N}{dV \; dm \; dc} = N_{tot} \times \frac{dP_{m}}{dm}(m) \times \frac{dP_{r}}{dV}(r)  \times \frac{dP_{c} }{dc}(c_{200}),
\end{equation}

where $P_m$ describes the probability of finding a subhalo with a given mass, $m$, $P_r$, the probability of finding a subhalo in a volume, $V$, and $P_{c}$, the scattering of concentration parameter, $c_{200}$. $N_{tot}$ represents the total number of substructures bounded to the main halo. The probability functions $P_r$, $P_m$ and $P_{c}$ are usually well described by statistical studies of subhalos resolved in sophisticated N-Body particle simulations of Milky Way-sized dark matter halos, for example Aquarius~\cite{Aquarius}, Via Lactea II \cite{vialacteaII} and ELVIS~\cite{elvis}.

 The mass distribution of subhalos in N-body simulations are well fitted by power-law functions, with small discrepancy of the spectral index $\alpha_{m}$ for different simulation sets. The power law fit of the differential mass abundance of Aquarius subhalos provides $\alpha_{m} = 1.9$, while ELVIS finds a slightly steeper value of $\alpha_{m} = 1.95$, and Via Lactea II finds $\alpha_{m} = 2.0$ for the same fit. Note that $\alpha_{m} = 2.00$ also provides a total predicted mass which is logarithmically divergent when extrapolated to small masses. However, even for the logarithmically divergent case, the total mass in substructures does not become large enough for this to happen, because a sharp cut-off in the subhalo mass spectrum is applied at a minimum mass. Interestingly, for $\alpha_{m} = 2.00$ the number of satellite galaxies above some minimal mass would increase roughly proportionally to the halo mass, i.e. $N_{\rm sat} \propto M_{\rm host}^{\alpha_{m} - 1}$. Moreover, recent studies~\cite{2011ApJ...736...59Z,2015ApJ...807..152S}  have found that the mean occupation number of satellites in observed galaxies up to $z \sim 1.2$ does favour values of $\alpha_{m} = 1.85 - 2.1$, as measured in simulations. 

Accretion dynamics over time shows that substructures are usually concentrated at distances closer to the galactic center. Because the 
the differential gamma-ray flux from subhalos is highly dependent on the distance of the source to the observer, the study of different parametrisations of d$P_r$/dV might provide insights into how the distribution of substructures around the main halo might impact the photon signal expected on Earth. In this study we examine three proposed distributions for d$P_r$/dV, as shown below. The first d$P_r$/dV parametrisation,
\begin{equation}
  \frac{dP_{r}}{dV}  = \rho_{0} \; exp \left[ - \frac{2}{\alpha_r} \left( \left(\frac{r}{r_{0}} \right)^{\alpha_r} - 1 \right)   \right],
  \label{eq::newtonr}
\end{equation}
named here N18, is given by reference~\cite{newton}, with $r_{0}$ = 43.0 kpc and $\alpha_r$ = 0.24. In this reference, a Bayesian inference method is employed to derive the total number of satellites in the galaxy based on two ingredients: the known census of Milky Way satellites and the radial distribution of subhalos, assumed to be the same obtained by the Aquarius set of simulations. The second parametrisation,
\begin{equation}
  \frac{dP_{r}}{dV} = \frac{\rho_{0}}{1 + \frac{r}{r_{0}}},
  \label{eq::hargisr}
\end{equation}
named H14, is given by reference~\cite{hargis}, with $r_{0}$ = 37.5 kpc. This study predicted the spatial distribution and total number of Milky Way satellites by completeness correcting the satellites detected by SDSS up to the Data Release (DR) 8. Because the completeness correction strategy requires a statistical description of the spatial distribution of satellites, H14 adopts the radial density of sub-halos given by the ELVIS simulations. For this reason, the slope of the mass distribution of satellites following the H14's spatial placement is assumed to be $\alpha_{m} = 1.95$. The same applies for N18, where an initial $\alpha_{m}$ = 1.90 has been chosen to characterise the mass abundance distribution of subhalos following this spatial distribution. 
The third parametrisation,
\begin{equation}
  \frac{dP_{r}}{dV}  = \frac{\rho_{0}}{ \frac{r}{r_{0}} \left( 1 + \frac{r}{r_{0}} \right)^{2} },
  \label{eq::koposovr}
\end{equation}
named here K8, is given by reference~\cite{koposov}, with $r_{0}$ = 16.7 kpc. K8 also adopts the completeness correction method, assuming, however, that sub-halos follow 
a Navarro-Frenk-White~\cite{nfw} (NFW) profile with $r_{s}$ = 10 kpc. Moreover, the list of known observed satellites used in the study contains only satellites observed by SDSS up to DR5. Because this spatial distribution parametrisation is not dependent on the results of N-body simulations, it lacks information on the parametrisation of the mass distribution. For this reason, in an attempt of generalisation, we assume that $\alpha_{m}$ = 1.90 in this scenario.

\begin{figure}[htb]
  \centering
  \includegraphics[scale=0.75]{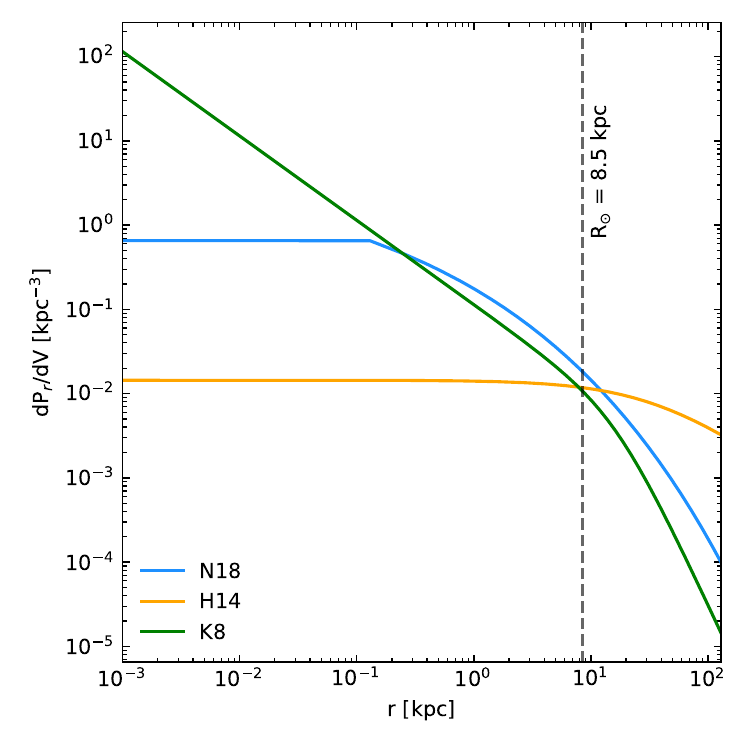}
  \caption{The radial distribution of subhalos ($\frac{dP_{r}}{dV}$) as a function of the distance from the galactic center ($r$) for models N18 (blue), H14 (orange) and K8 (green). The distance of the Earth to the Milky Way center \cite{earth} is represented by the dashed line.}
  \label{fig:dpdv}
\end{figure}

In all parametrisations, the normalisation factor, $\rho_{0}$, is such that $\frac{dP_{r}}{dV}$ is normalized to one, just like expected for a probability density function. The behaviour of N18, H14 and K8 distributions as a function of the distance from the galactic center is shown in figure~\ref{fig:dpdv}.

The total number of substructues expected by each study N18, H14 and K8 is shown in table~\ref{tab:tablenumbersubs}. For N18 we present the mean value, upper and lower limits of the 68 $\%$ C. L. of the expected total population of subhalos. For H14 we present only the upper and lower limits of the 90 $\%$ C. L. For K8 we use the luminosity function given by reference~\cite{koposov} to calculate the total number of subhalos between absolute magnitude $M_{V}$ = -20 and $M_{V}$ = 0. The mass range [$m_{min}$, $m_{max}$] is set by the resolution of the Aquarius (assumed for N18) or ELVIS (assumed for H14) set of simulations and by the value of $m_{max}$ such that $\int_{m_{max}}^{M_{vir}} \frac{dN_{m}}{dm}(m) \; dm$ < 1, where $M_{vir}$ is the MW virial mass. Because model K8 is not based on results of dark matter simulations, it receives the same inputs as N18 for the characterisation of the mass distribution.

A third ingredient necessary to characterise the inner shapes of subhalos is the concentration parameter \cite{diemer}, $c_{200}$, defined as the ratio of the subhalo virial radius and $r_{-2}$, the radius at which the slope of the logarithmic of the dark matter density profile is -2, i.e., d$log(\rho)$/d$log(r)$|$_{r_{-2}}$ = -2. In practice, halo concentrations are determined via N-body simulations, by fitting the dark matter distribution in galaxy halos. This parameter has been extensively studied, with several analytical approximations being proposed in the literature \cite{bullock} and a universal parametrisation at z = 0 being proposed by \cite{sanchez}. Given the mass of a sub-halo, or its virial radius, and the concentration parameter, the dark matter density profile of substructures is completely delineated~\cite{moline}. 

In an attempt to consider the effects of subhalo properties on their dark matter signal, we investigate two parametrisations for $c_{200}$. The first one, given in reference~\cite{moline} and named here MO17, is obtained via the study of subhalo properties, for example mass and distance to the halo center, which are derived from statistics of Via Lactea II (VL-II)~\cite{vialacteaII} and ELVIS simulations. 

\begin{figure}[htb]
  \centering
  \includegraphics[scale=0.7]{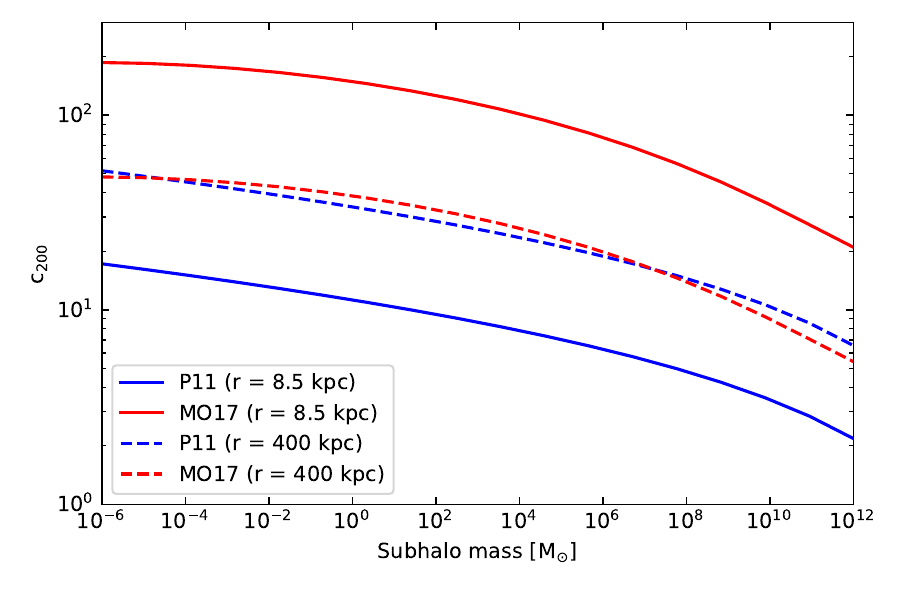}
  \caption{The concentration parameter, $c_{200}$, for MO17 (blue) and P11 (red) models as a function of subhalo mass at a distance of 8.5 kpc (solid line) and 400 kpc (dashed line) from the host halo.}
  \label{fig:c200}
\end{figure}

The second model, given in reference~\cite{pieri} and named here P11, is obtained from statistical studies of subhalos resolved in the Aquarius simulations. Figure~\ref{fig:c200} shows the behaviour of MO17 and P11 as a function of the subhalo mass at different distances from the galactic center. Note that, in general, P11 provides lower concentration values in respect MO17 at the same distance from the host. This effect gets lower as the distance increases. Because simulations show a halo-to-halo scatter of the concentration parameter~\cite{bullock}, $c_{200}$ is assumed to be log-normal distributed around the mean $\overline{c_{200}}$, with dispersion $\sigma_{200}$. P11 finds $\sigma_{200}$ = 0.14 for Aquarius subhalos, while  MO17 computes a typical scatter is $\sigma_{200}$ = 0.11 for VL-II and
$\sigma_{200}$ = 0.13 for the ELVIS simulations. 

\renewcommand{\arraystretch}{1.8}
\begin{table}[h!]
  \centering
  \begin{tabular}{|c|cc|} \hline 
  Model  &  $N_{tot} (< R_{vir}$) & [$m_{min}$, $m_{max}$] ($M_{\odot}$, $M_{\odot}$)\\ 
  \hline
    N18 (68 $\%$ m. C. L.) & 115 & [2 $\times$ 10$^{5}$, 1 $\times$ 10$^{11}$]\\
    N18 (68 $\%$ u. l. C. L.) & 151 & [2 $\times$ 10$^{5}$, 1 $\times$ 10$^{11}$] \\
    N18 (68 $\%$ l. l. C. L.) & 90 & [2 $\times$ 10$^{5}$, 1 $\times$ 10$^{11}$] \\
    H14 (90 $\%$ u. l. C. L.) & 896 & [2.35 $\times$ 10$^{4}$, 2.14 $\times$ 10$^{9}$]  \\ 
    H14 (90 $\%$ l. l. C. L.) & 168 & [2.35 $\times$ 10$^{4}$, 2.14 $\times$ 10$^{9}$] \\ 
    K8  ($-20 < M_{V} < 0$) & 136 & [2 $\times$ 10$^{5}$, 1 $\times$ 10$^{11}$] \\
    \hline
  \end{tabular}
  \caption{Total number of subhalos $N_{tot}$ expected within the MW virial radius ($R_{vir}$) by population studies and respective mass range for each radial model.}
  \label{tab:tablenumbersubs}
\end{table}
\renewcommand{\arraystretch}{1}

With the characterisation of the distribution of subhalo populations complete, we can investigate the expected gamma-ray flux from dark matter annihilation in MW substructures. The differential photon flux that reaches an observer and that is emitted by a dark matter source is given by
\begin{equation}
  \frac{d\phi_{\gamma}}{dE} = \frac{\braket{\sigma v}}{8 \pi m_{\chi}^{2}} \times \frac{d N _{f}}{dE} \times B_{f} \times J \,\, ,
  \label{eq::diff-flux}
\end{equation}
where $\braket{\sigma v}$ is the velocity averaged annihilation cross-section, $dN_{f}/dE$ is the energy spectrum of the gamma-rays produced via annihilation in channel $f$, $B_{f}$ = $\braket{\sigma v}_{f}$/$\braket{\sigma v}$ is the specific branching ratio and $m_{\chi}$ is the dark matter particle mass. In this paper, we analyse the $\tau^{+}\tau^{-}$ and $b\bar{b}$ annihilation channels, assuming $B_{f} = 1$ for each case.

The astrophysical J-factor,
\begin{equation}
  J = \int_{\Delta \Omega} \int_{l.o.s.} \rho (l) ^{2}  \,\, d\Omega  \,\,\, dl \,\, ,
\end{equation}
is the integral in the solid angle, $\Delta \Omega$, and in the line of sight (l.o.s.) of the dark matter density squared, $\rho(l)^{2}$. 

For the inner profile of the distribution of dark matter in each subhalo, we investigate the effect of two traditional parametrisations, Einasto~\cite{einasto} (Ein) and NFW:
\begin{eqnarray}
  \rho_{\mathrm{Ein}} & = & \rho_s \exp{\left[ -\frac{2}{0.17}\left( \frac{r}{r_s} \right)^{0.17} \right]},
\end{eqnarray}
  
\begin{eqnarray}  
  \rho_{\mathrm{NFW}} & = &\frac{\rho_s}{ \left( \frac{r}{r_s} \right) \left( 1 + \frac{r}{r_s} \right)^2 } \, \, .
\end{eqnarray}

With all the ingredients set, and in order to study the effect of subhalos properties on dark matter annihilation, we can derive global substructure models by combining a) the subhalos inner profile, b) the radial distribution parametrisations (d$P_r$/dV), c) the mass abundance distribution (d$P_m$/dm), d) the total number of subhalos and e) the concentration parameter ($c_{200}$). Different distributions and parameters will affect the total mass of the galaxy and its substructures, as well as their expected dark matter signal. Besides, there are some environmental effects (e.g. subhaloes are expected to be disrupted by tidal forces if baryons are present) which may generate a correlation between the spatial and mass distributions~\cite{Stref:2016uzb}. However in the DM-only simulations that we are using here, there is little correlation between spatial, mass and $c_{200}$ distributions~\cite{Aquarius,pieri}, so in principle, a permutation of the different parametrizations is physically allowed. For instance, although d$P_r$/dV parametrisations are based on different simulations, which found different values of $\alpha_{m}$, these are just the central fitted median values of a measured distribution with typical standard deviation of $\pm 0.3$~\cite{vialactea,Aquarius}. Finally, results from simulations also show that the subhaloes radial distribution does necessarily follow the inner density profile of the main halo and subhaloes~\cite{Aquarius,pieri,Stref:2016uzb}, so we also test the impacts of this possibility here. 

However, due to computational limitations, it is not feasible to perform the analysis of the entire possible range of parameter permutations. For this reason, we choose N18 as a default radial distribution model, in subhalo model 1 (SM1), and limit the study of H14 and K8 distributions to models SM8, 9 and 10. In the same direction, we must define a default option for the remaining parameters. Those are: the Einasto profile for the inner dark matter content of subhalos, the mean value of the 68 $\%$ C. L. of the total number of subhalos expected by N18, $\alpha_{m}$ = 1.9 for the slope of the mass distribution and MO17 with dispersion $\sigma_{200}$ = 0.13 for the concentration parameter. Table~\ref{tab:models} shows a summary of the ten subhalo models that are considered in this study.

Model SM1 tests only the default options for each parameter, while all remaining models test only the effect of one alternative value of a subhalo parameter,  setting all other ingredients to default. In summary, we limit to the study of a modification to a NFW profile for the inner dark matter content of subhalos in model 2, the logarithmically divergent $\alpha_{m}$ = 2.0 in model 3, the alternative value of $\sigma_{200}$ = 0.11 found by MO17 for VL-II subhalos in model 4, the upper and lower values of N18's expected total number of subhalos in models 5 and 6, and P11 as a secondary choice for $c_{200}$ in model 7. The radial distribution given by K8, and, consequentially, its expectation for $N_{tot}$, is tested in model SM8. For H14, we test the upper and lower values of $N_{tot}$ (90 \% C. L.) in models SM9 and SM10, respectively.

\renewcommand{\arraystretch}{1.8}
\begin{table}[h]
  \centering
  \resizebox{\linewidth}{!}{
    \begin{tabular}{|c|cccccccccc|} \hline \toprule
      Parameter & SM1 & SM2 & SM3 & SM4 & SM5 & SM6 & SM7 & SM8 & SM9 & SM10 \\ \midrule \midrule
      \hline
      inn. prof. & Ein & NFW & Ein & Ein & Ein & Ein & Ein  & Ein & Ein & Ein \\ \midrule
      spat. dist & N18 & N18 & N18 & N18 & N18 & N18 & N18 & K8 & H14 & H14 \\ \midrule
      $\alpha_{m}$ & 1.9 & 1.9 & 2.0 & 1.9 & 1.9 & 1.9 & 1.9 & 1.9 & 1.95 & 1.95 \\ \midrule
      $N_{calib}$ & 115 & 115 & 115 &  115 & 151 & 90 & 115 & 136 & 896  & 168 \\ \midrule
      $c_{200}$ & MO17 & MO17 & MO17 & MO17 & MO17 & MO17 & P11 & MO17 & MO17 & MO17 \\ \midrule
      $\sigma_{200}$ & 0.13 & 0.13  & 0.13 & 0.11 & 0.13 & 0.13 & 0.14  & 0.13 & 0.13  & 0.13 \\  \hline
  \end{tabular} }
  \caption{Summary of the set of parameters for each model. The varied criteria are: the inner profiles of substructures, the radial distribution, the spectral index of the power law that describes the mass distribution ($\alpha_{m}$), the calibration number of substructures ($N_{calib}$), the concentration relation ($c_{200}$) and the scatter of the concentration relation ($\sigma_{200}$).}%
  \label{tab:models}
\end{table}
\renewcommand{\arraystretch}{1}
%


\section{Simulation of the sources and CTA detection}
\label{sec:sim}

\begin{figure}[h]
  \centering
    \includegraphics[scale=0.55]{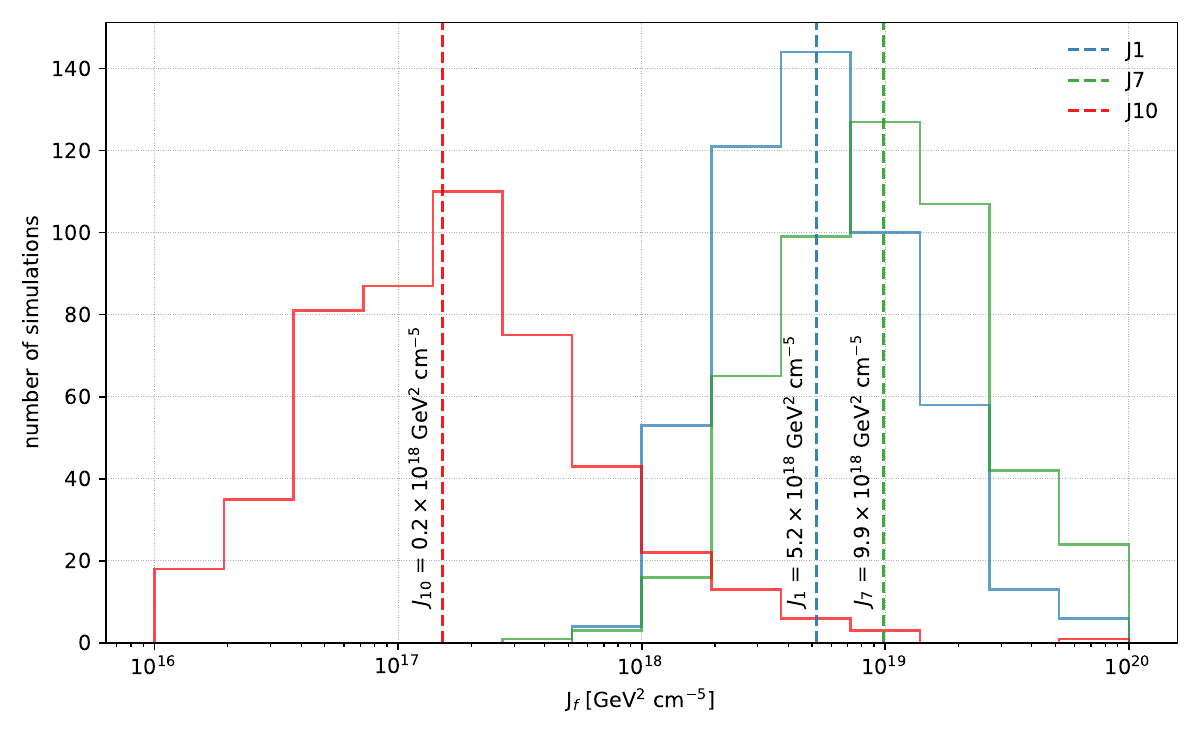}
  \caption{Distribution of simulated J-factors for models SM1, the standard choice for comparison, SM7 and SM10, which provided the highest and lowest median J-factors, respectively. The dashed line shows the median J-factor for each model.}
\label{fig:Jfcompare}
\end{figure}

For each model presented in table~\ref{tab:models}, we calculated 500 sky-maps with the \texttt{CLUMPY} code~\cite{clumpy3}. In each realisation, the subhalo with the highest $J_{f}$ is was selected. This guarantees the study of the subhalo population with the highest possible gamma-ray signal. The distribution of the 500 highest $J_{f}^{i}$ from models SM$_{i}$ $i \in \{ 1, 7, 10 \}$, i.e., the default and models with the highest and lowest median J-factors, is shown in Figure~\ref{fig:Jfcompare}.

Table~\ref{tab:statistics:models} shows the statistical parameters of the J-factor distribution for each SM$_{i}$ model: mean, median, and upper and lower limits of the 95 \% C. L. It also shows the mean distance to the observer ($\overline{d_{obs}}$) and virial mass ($\overline{m_{200}}$). From this table, it is possible to infer the general effects of the SM models in the J-factor calculation:

\begin{itemize}
\item NFW generates smaller J-factors than Einasto profile when fixing the remaining parameters. The difference can reach up to 50\%. This can be explained by the fact that the integration of the Einasto parametrization over the extent of a subhalo of any size yields higher dark matter masses in comparison to NFW when assuming the same scale parameters;
\item The radial distribution of subhalos has a moderate effect on J-factor values. H14 reduces J-factor by 15\%, regarding N18 and K8. This might be motivated by the fact that a flatter behaviour of d$P_r$/dV, as seen for H14, results in N18 and K8 resolving more subhalos at smaller distances from the observer, which also strengthens the gamma-ray signal compared to H14;
\item A steeper value of $\alpha_{m}$ disfavours the generation of higher mass subhalos in comparison to milder spectral indexes, which in turn results in the production of smaller J-factors. This can be seen in model SM2, where the mean J-factor is 5\% lower in respect to SM1.
\item  The concentration parameter model has a large effect on J-factor values. P11 parametrisation results in J-factors twice as large as the MO17 paremetrisation. This can be explained by the fact that P11 provides lower values of $c_{200}$ in comparison to MO17 at the same subhalo distance to the host center, which in turn supplies higher densities for the dark matter distribution within the subhalo, and, therefore, strengthens the gamma-ray signal;
\item The J-factor is not very sensitive to the dispersion values tested in this work. The maximum variation from $\sigma_{200} = 0.11$ to $\sigma_{200} = 0.13$ is 5\%.
\end{itemize}


In the next sections, we are going to report the results on the detectability by CTA of a gamma-ray signal from DM annihilation from subhalos using the simulated J-factors. Our intention is to explore the extremes allowed by the models, therefore we will show results for two limits, corresponding to the  5\% of the J-Factor distribution (low signal) and 95\% of the J-Factor distribution (high signal). We name these extremes Faint and Bright sources for clarity. These hypothetical sources represent the lower and upper limits of the 95\% C. L. among the 500 highest $J_{f}^{i}$ for each model SM$_{i}$.

\renewcommand{\arraystretch}{1.8}

\begin{table}[h]
  \resizebox{\linewidth}{!}{
    \begin{tabular}{|c|cccccccccc|} \toprule
    \hline
      Property & SM1 & SM2 & SM3 & SM4 & SM5 & SM6 & SM7 & SM8 & SM9 & SM10 \\ \hline \midrule \midrule
      $\overline{d_{obs}}$ [kpc] &  11.1 & 11.6 & 10.5 & 11.0 & 11.0 & 11.9 & 11.0  & 15.6 & 29.6 & 41.9 \\ \midrule
      $\overline{m_{200}}$ [10$^{7}$ M$_{\odot}$] &  4.6 & 8.8 & 6.6 & 6.0 & 6.9 & 7.2 & 5.9 & 14.3 & 12.3 & 2.2  \\ \midrule
      log $\left( \frac{\overline{J_{f}}}{10^{18} \; \mathrm{GeV^{2}/cm^{5}} } \right)$ & 8.4 & 3.8 & 8.0 & 13.0 & 10.1 & 6.8 & 20.7 & 5.0 & 2.1 & 0.6 \\ \midrule
      log $\left( \frac{\widehat{\mathrm{J}}_{f}}{10^{18} \; \mathrm{GeV^{2}/cm^{5}} } \right)$ & 5.2 & 2.5 & 4.4 & 5.0 &  6.3 & 3.8 & 9.9 & 3.1 & 0.8 & 0.2 \\ \midrule
      log $\left( \frac{ 95 \% \; \mathrm{u. l.} \; \mathrm{C. L.} \; \mathrm{J}_{f} }{10^{19} \; \mathrm{GeV^{2}/cm^{5}} } \right)$  &  2.2  & 1.1 & 2.6 & 2.9 & 2.8 & 1.9 & 6.8 & 1.5 & 0.6 & 0.2  \\ \midrule
      log $\left( \frac{ 95 \% \; \mathrm{l. l.}  \; \mathrm{C. L.} \; \mathrm{J}_{f} }{10^{18} \; \mathrm{GeV^{2}/cm^{5}} } \right)$  & 1.4 & 0.8 & 1.3 & 1.6 & 1.8 & 1.0 & 2.1 & 1.1 & 0.2 & 0.02 \\ \bottomrule
      \hline
  \end{tabular}}
  \caption{Mean properties of subhalos resolved for each model SM$_{i}$ $i \in \{1, ..., 10 \}$: distance to the observer ($\overline{d_{obs}}$), virial mass ($\overline{m_{200}}$) and J-factor ($\overline{J_{f}}$). Table also brings the median ($\widehat{J_{f}}$), 95 $\%$ C. L. upper limit (95 $\%$ C. L. u. l. J$_{f}$) and 95 $\%$ C. L. lower limit (95 $\%$ C. L. l. l. J$_{f}$) J-factors among the set.}
  \label{tab:statistics:models}
\end{table}
\renewcommand{\arraystretch}{1}


\section{Results}
\label{sec:results}

\subsection{Simulation of the detection by CTA}
\label{sec:cta:detection}

We use the CTOOLs 1.6.3 package~\cite{ctoolsweb, ctools} to simulate how CTA would measure a source signal. The instrument response functions (IRFs) of CTA~\cite{2019APh...111...35A} with calibration database prod3b-v2 were used in the calculation. The sources were positioned in the Southern and Northern hemispheres at right ascension and declination (180$^\circ$, -45$^\circ$) (J2000) and (180$^\circ$, 45$^\circ$) (J2000), respectively. This choice of declination guarantees that the source is visible over a wide range of elevation angles, i.e from 25$^{\circ}$ to 85$^{\circ}$, allowing the total observation time to be split up over the span of several days and also guarantying that data are taken in the regime of CTA's lowest energy threshold, which is particularly important at low energies, i.e, $\sim$ 20 GeV and $\sim$ 150 GeV for the north and south arrays, respectively. Moreover, this position of the source allows for its observability above elevation > 30$^{\circ}$ for more than 600 hours during dark time, which accommodates other science cases with observation time of hundreds or dozens of hours, or which require high source elevations. For similar reasons, and given that simulations of CTA observations with CTOOLs require the input of a IRF parameter file, we have chosen IRFs computed for zenith 40$^{\circ}$. This choice of zenith, among others available for prod3b-v2 (20$^{\circ}$ and 60$^{\circ}$) does not provide a critical impact on the sensitivity of the simulations, since the lowest value of $m_{\chi}$ tested in this work is in the TeV regime. We simulated observations of 50 hours. CTOOLs simulate the signal detected by CTA given a spatial and spectral model of the subhalo. It also simulates the background rate following the parametrisations in the IRF.

A maximum likelihood search is implemented to test the compatibility between the simulated data and a subhalo source model. Given a model hypothesis, $\mathcal{M}$, the likelihood ratio is
\begin{equation}
  \lambda = \frac{\mathcal{L}(\mathcal{M}_{bkg}(\mathbf{\Theta}_{bkg})|\mathbf{X}  )}{ \mathcal{L}(\mathcal{M}_{bkg}(\mathbf{\Theta}_{bkg}) + \mathcal{M}_{sig}(\mathbf{\Theta}_{sig})|\mathbf{X}  )},
  \label{likelihood}
\end{equation}
where $\textbf{X} = (N_{obs}, E’, \mathbf{p’}$) stores the simulated number of events, reconstructed energy and reconstructed incident direction of the photons. $\mathbf{\Theta}$ contains the adjusted free parameters that maximise the likelihood function $\mathcal{L}$. The function $\lambda$ tests whether the probability of only background events, known as alternative hypothesis, is statistically preferable to the null hypothesis, i.e., the presence of background and signal events. In this case, $\mathbf{\Theta}_{sig} = \braket{\sigma v}$. If $\lambda$ is maximized, the output of the fitting procedure is the value of $\braket{\sigma v}$ which best describes the simulated data. The Test Statistics, TS, of a likelihood $\lambda$,
\begin{equation}
  TS = -2 \; log \lambda,
\end{equation}
is used to reject the alternative hypothesis at a given confidence level.

\subsection{Calculation of $\braket{\sigma v}$ in case of source detection with the CTA}
\label{sec::detec}

\begin{figure}
  \centering
  \begin{subfigure}[b]{0.48\textwidth}
    \centering
    \includegraphics[width=\textwidth]{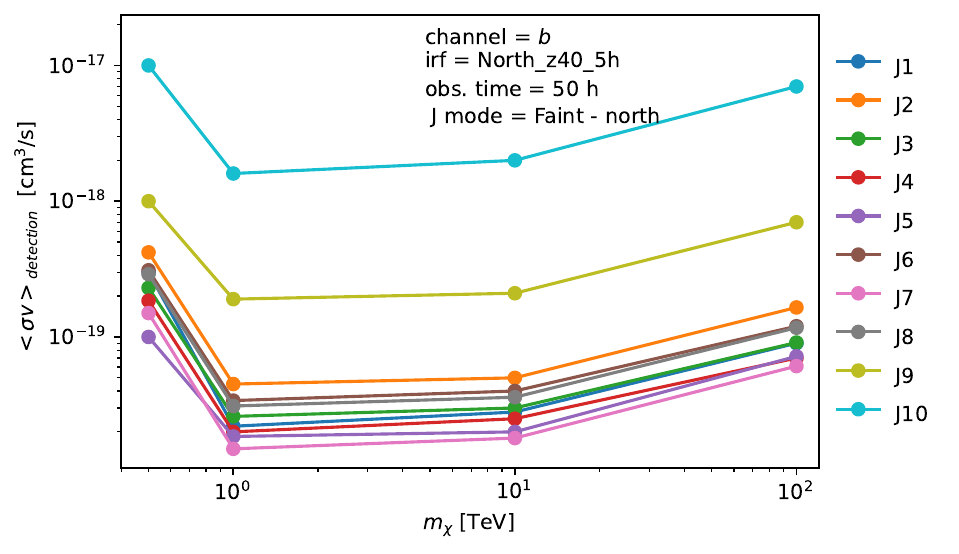}
    \caption{Faint --- North --- $b\bar{b}$}
    \label{fig:n:b}
  \end{subfigure}
  \hfill
  \begin{subfigure}[b]{0.48\textwidth}
    \centering
    \includegraphics[width=\textwidth]{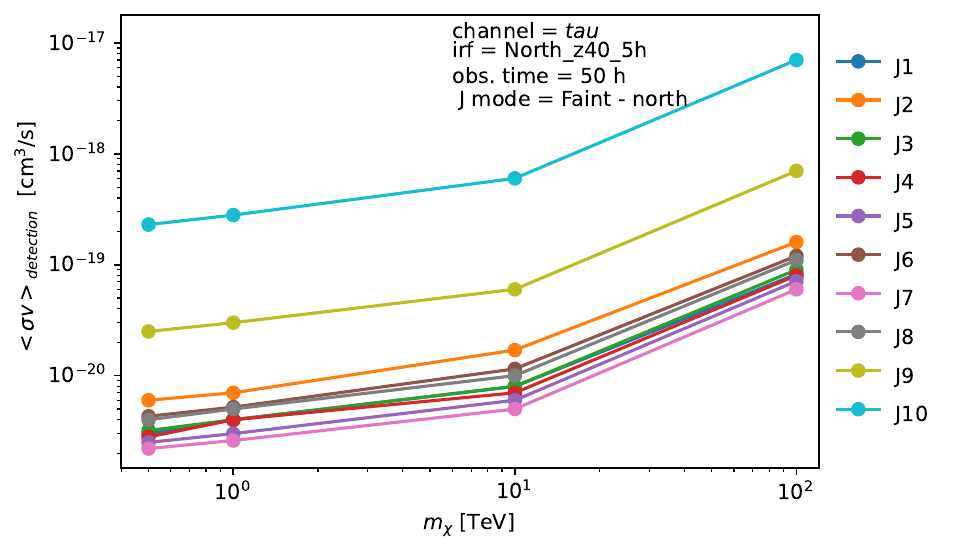}
    \caption{Faint --- North --- $\tau^{+}\tau^{-}$}
    \label{fig:n:t}
  \end{subfigure}

   \centering
  \begin{subfigure}[b]{0.48\textwidth}
    \centering
    \includegraphics[width=\textwidth]{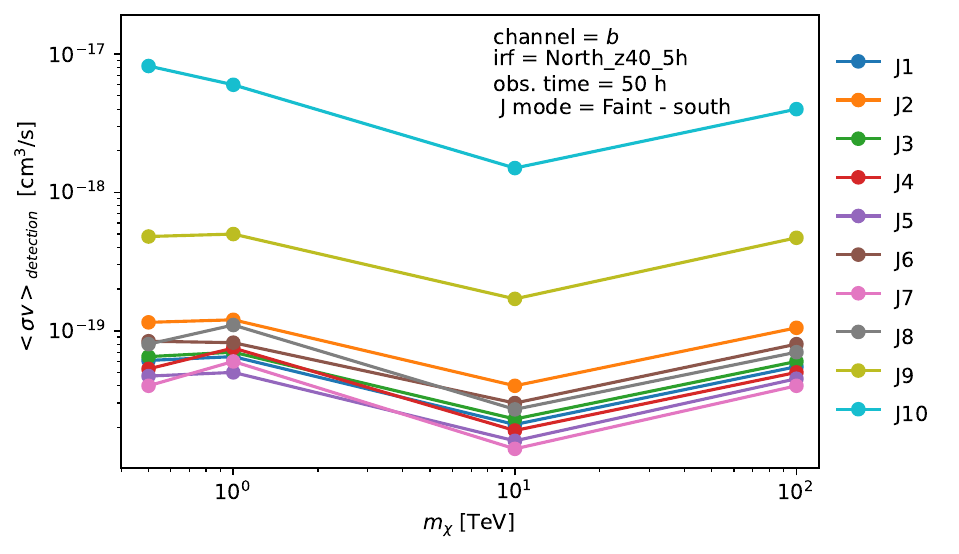}
    \caption{Faint --- South --- $b\bar{b}$}
    \label{fig:n:b}
  \end{subfigure}
  \hfill
  \begin{subfigure}[b]{0.48\textwidth}
    \centering
    \includegraphics[width=\textwidth]{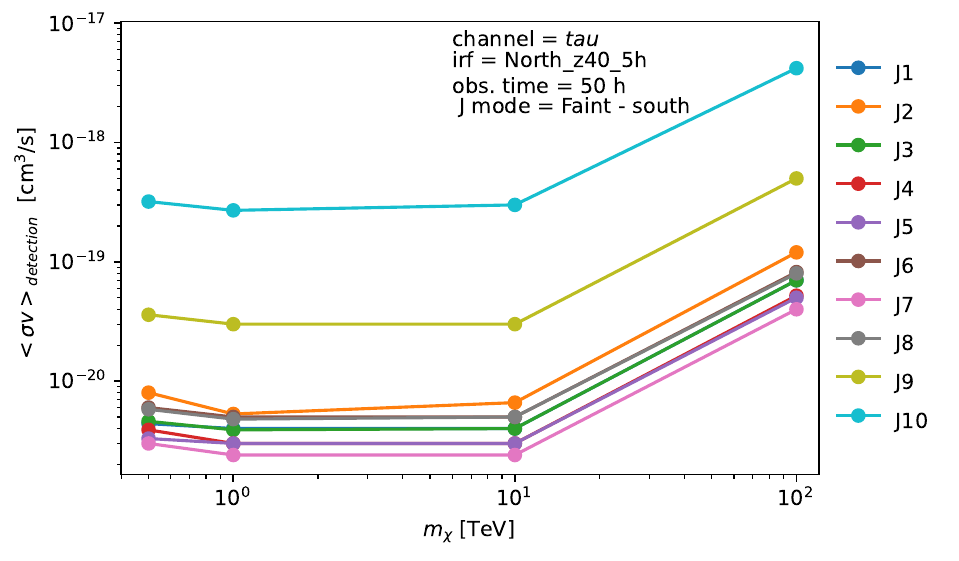}
    \caption{Faint --- South --- $\tau^{+}\tau^{-}$}
    \label{fig:n:t}
  \end{subfigure}

 \centering
  \begin{subfigure}[b]{0.48\textwidth}
    \centering
    \includegraphics[width=\textwidth]{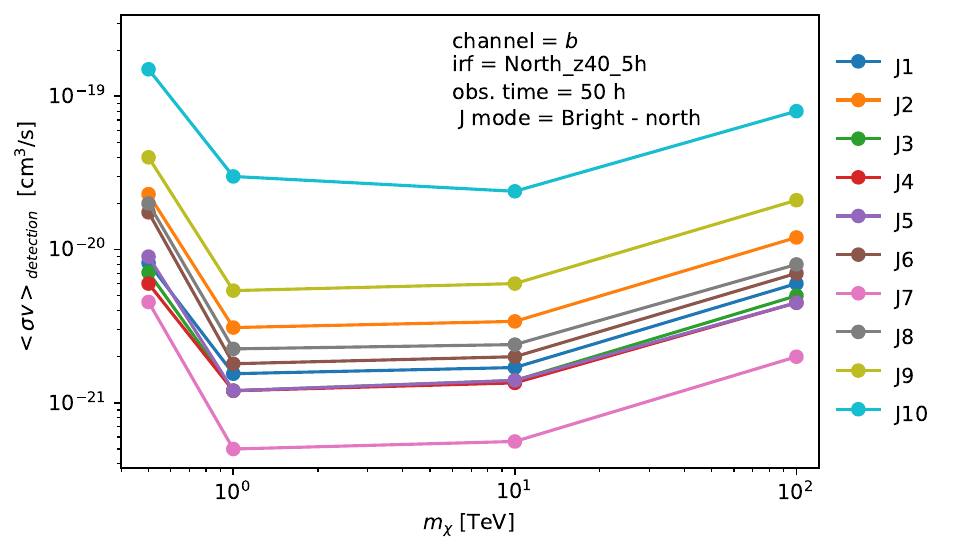}
    \caption{Bright --- North --- $b\bar{b}$}
    \label{fig:n:b}
  \end{subfigure}
  \hfill
  \begin{subfigure}[b]{0.48\textwidth}
    \centering
    \includegraphics[width=\textwidth]{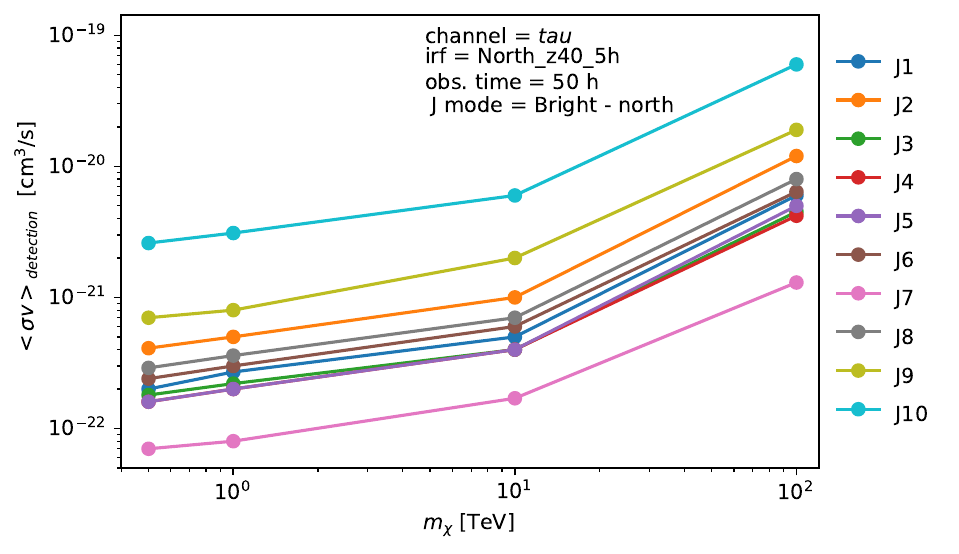}
    \caption{Bright --- North --- $\tau^{+}\tau^{-}$}
    \label{fig:n:t}
  \end{subfigure}

   \centering
  \begin{subfigure}[b]{0.48\textwidth}
    \centering
    \includegraphics[width=\textwidth]{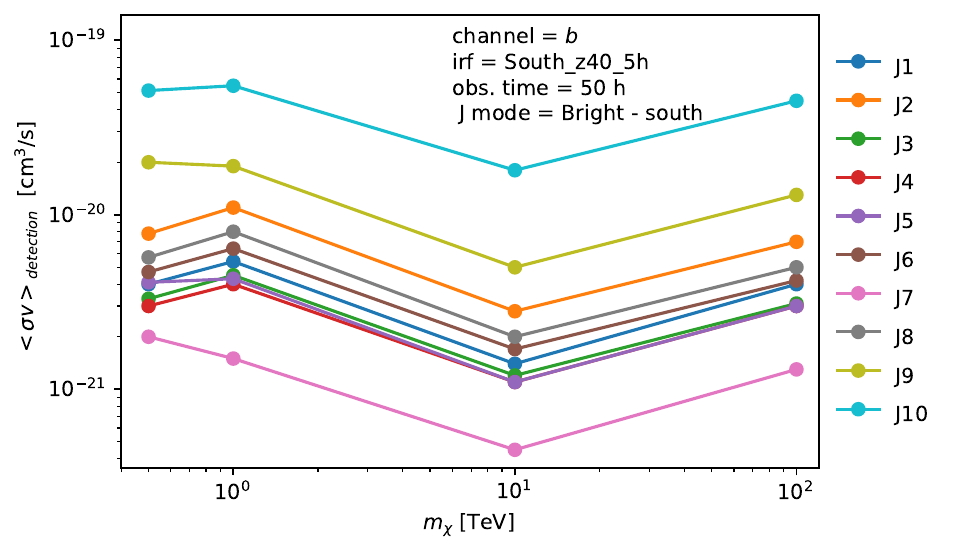}
    \caption{Bright --- South --- $b\bar{b}$}
    \label{fig:n:b}
  \end{subfigure}
  \hfill
  \begin{subfigure}[b]{0.48\textwidth}
    \centering
    \includegraphics[width=\textwidth]{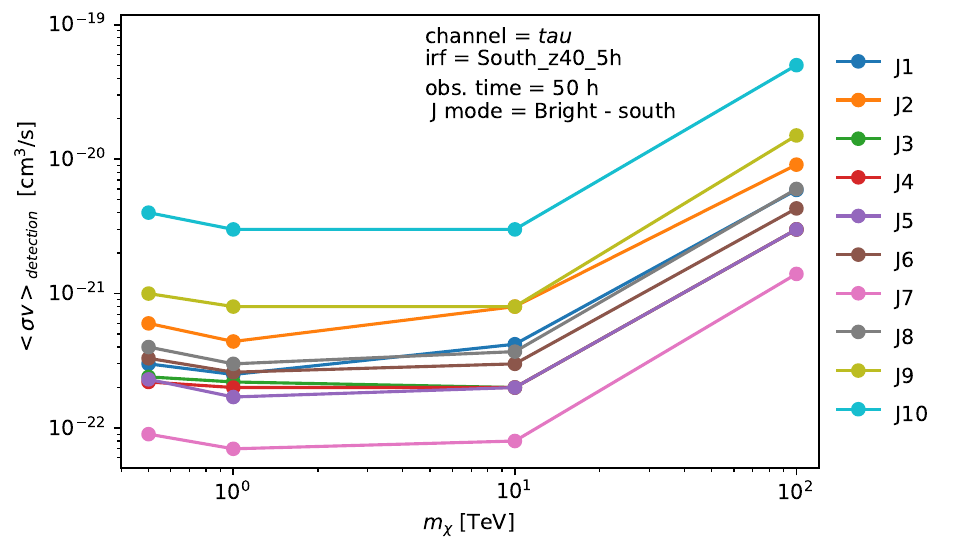}
    \caption{Bright --- South --- $\tau^{+}\tau^{-}$}
    \label{fig:n:t}
  \end{subfigure}
  \caption{CTA 5$\sigma$ detection sensitivity on the velocity-weighted cross-section $\braket{\sigma v}$ for different source strengths, array site and annihilation channel.}
  \label{fig:detection}
\end{figure}

We search the value of $\braket{\sigma v}$ for which $TS = 25$, which is equivalent to a 5$\sigma$ signal detection by CTA. The calculation is repeated for $m_{\chi} =$ 0.5, 1.0, 10.0, and 100 TeV. Figure~\ref{fig:detection} shows the sensitivity curves. For all mass points except at $m_{\chi}$ = 100.0 TeV, the $\tau^{+}\tau^{-}$ channel provided the most constraining sensitivity curves. When comparing both hemispheres, the sensitivity curves are essentially equivalent, with ratio reaching less than one order of magnitude. 

As expected by the obtained J-factors in the previous section, model SM7 offers the most constrained sensitivity curves for both hemispheres and annihilation channels, except at $m_\chi =$ 0.5 TeV for a faint source at the south hemisphere, where SM5 provides a more constrained $\braket{\sigma v}$  (5$\sigma$). No model offers the possibility of probing the thermal values of $\braket{\sigma v}$. Also as expected, due to the lower simulated J-factors, model SM10 provides the least constraining sensitivity curves. When comparing both hemispheres, the highest ratio between a $\braket{\sigma v} \times m_{\chi}$ point for the south and north hemisphere was $\sim$ 1.05, for a Bright SM3 source and annihilation on the $b\bar{b}$ channel.

\subsection{Calculation of the upper limit on $\braket{\sigma v}$ in case of no detection with the CTA}
\label{sec::nodet}

\begin{figure}
  \centering
  \begin{subfigure}[b]{0.48\textwidth}
    \centering
    \includegraphics[width=\textwidth]{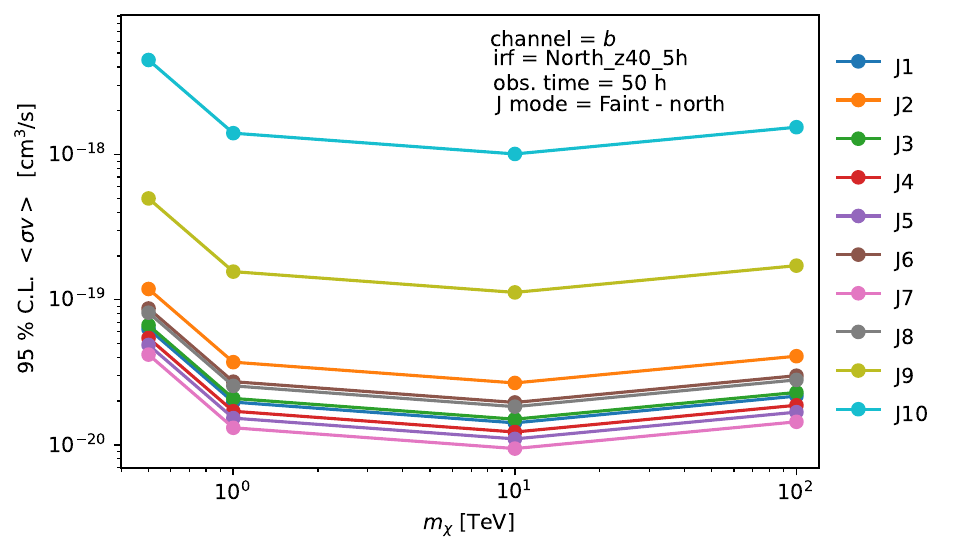}
    \caption{Faint --- North --- $b\bar{b}$}
    \label{fig:n:b}
  \end{subfigure}
  \hfill
  \begin{subfigure}[b]{0.48\textwidth}
    \centering
    \includegraphics[width=\textwidth]{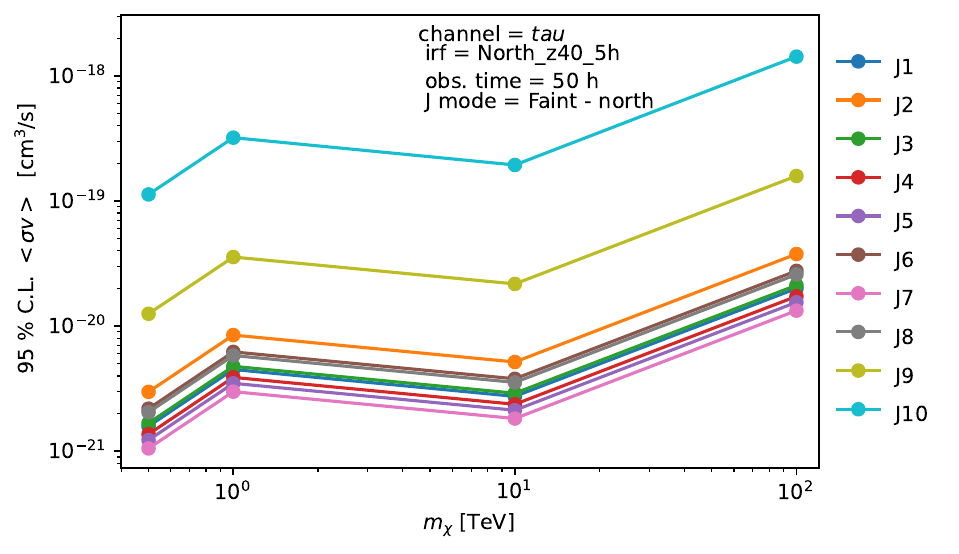}
    \caption{Faint --- North --- $\tau^{+}\tau^{-}$}
    \label{fig:n:t}
  \end{subfigure}

   \centering
  \begin{subfigure}[b]{0.48\textwidth}
    \centering
    \includegraphics[width=\textwidth]{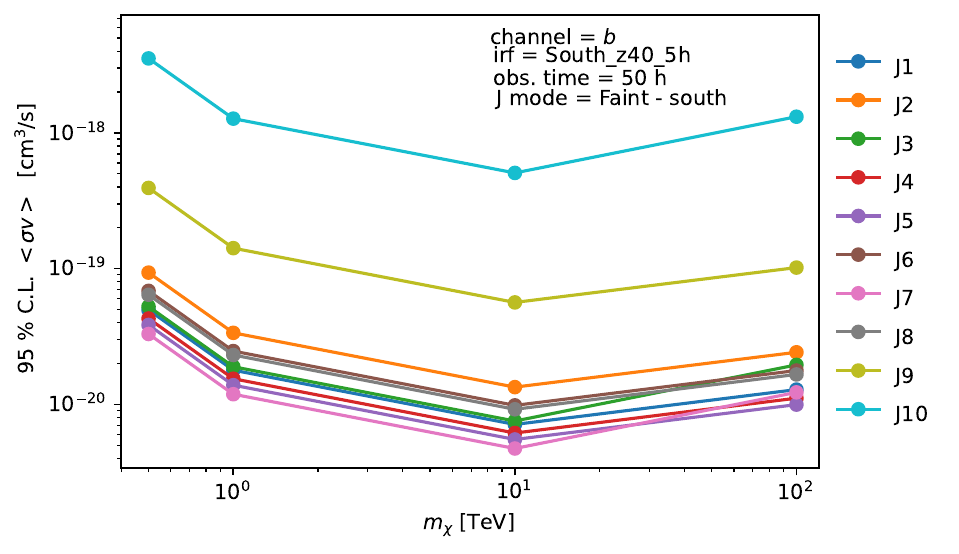}
    \caption{Faint --- South --- $b\bar{b}$}
    \label{fig:n:b}
  \end{subfigure}
  \hfill
  \begin{subfigure}[b]{0.48\textwidth}
    \centering
    \includegraphics[width=\textwidth]{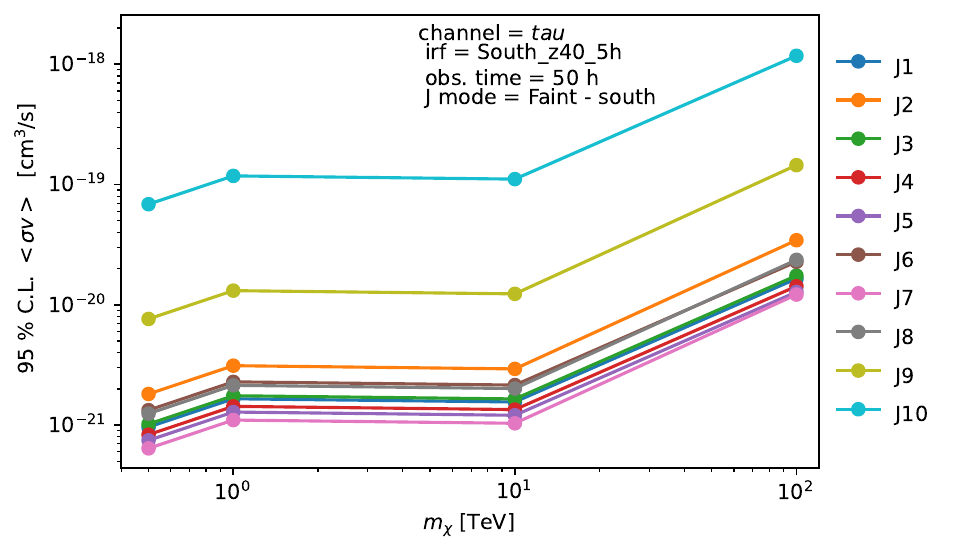}
    \caption{Faint --- South --- $\tau^{+}\tau^{-}$}
    \label{fig:n:t}
  \end{subfigure}

 \centering
  \begin{subfigure}[b]{0.48\textwidth}
    \centering
    \includegraphics[width=\textwidth]{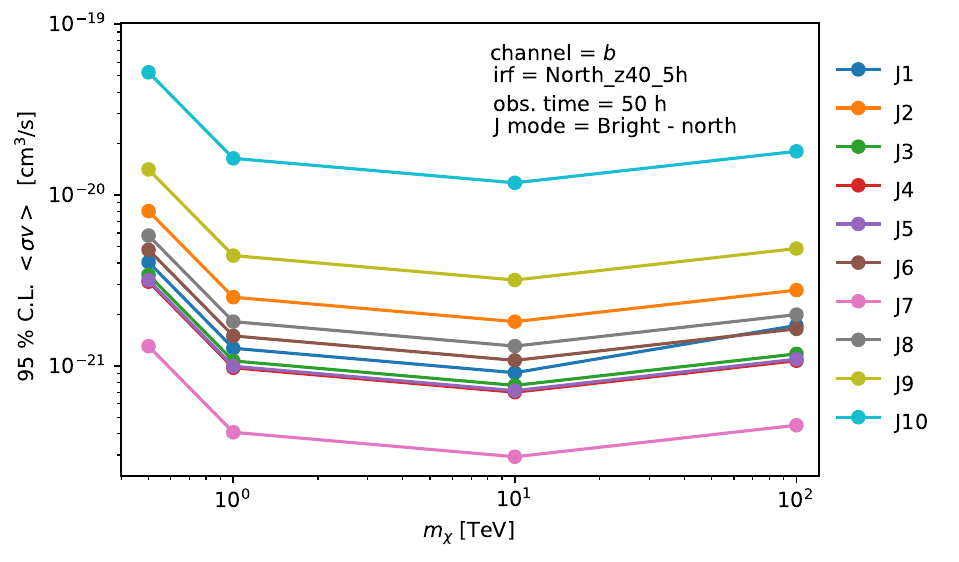}
    \caption{Bright --- North --- $b\bar{b}$}
    \label{fig:n:b}
  \end{subfigure}
  \hfill
  \begin{subfigure}[b]{0.48\textwidth}
    \centering
    \includegraphics[width=\textwidth]{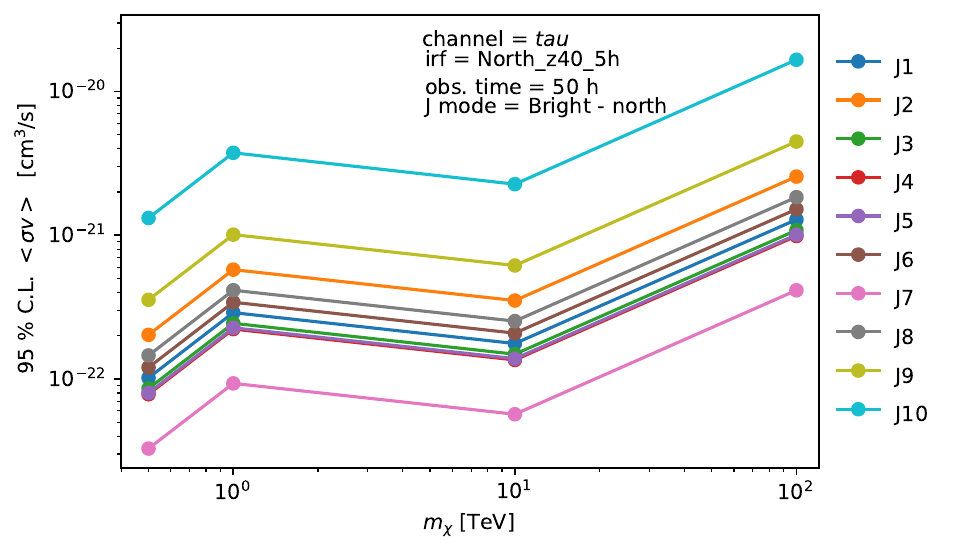}
    \caption{Bright --- North --- $\tau^{+}\tau^{-}$}
    \label{fig:n:t}
  \end{subfigure}

   \centering
  \begin{subfigure}[b]{0.48\textwidth}
    \centering
    \includegraphics[width=\textwidth]{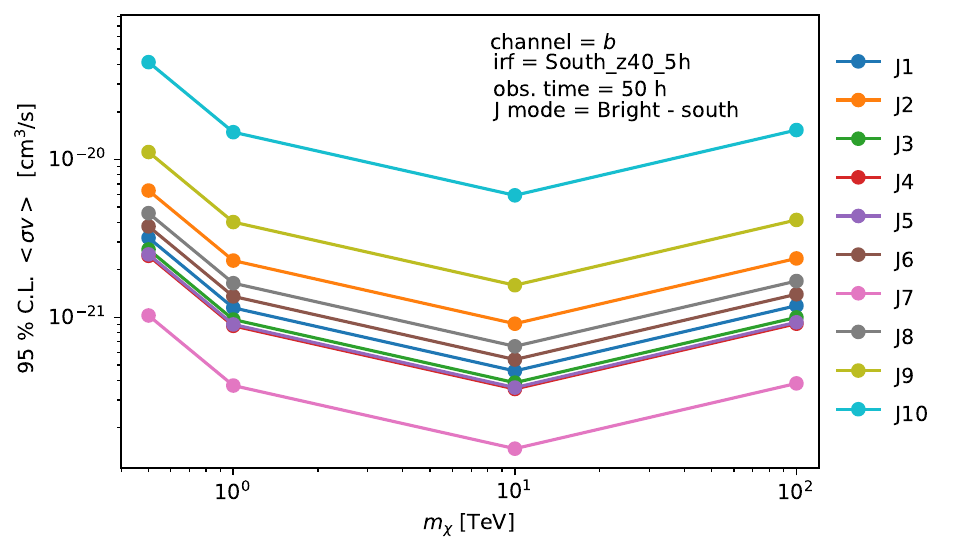}
    \caption{Bright --- South --- $b\bar{b}$}
    \label{fig:n:b}
  \end{subfigure}
  \hfill
  \begin{subfigure}[b]{0.48\textwidth}
    \centering
    \includegraphics[width=\textwidth]{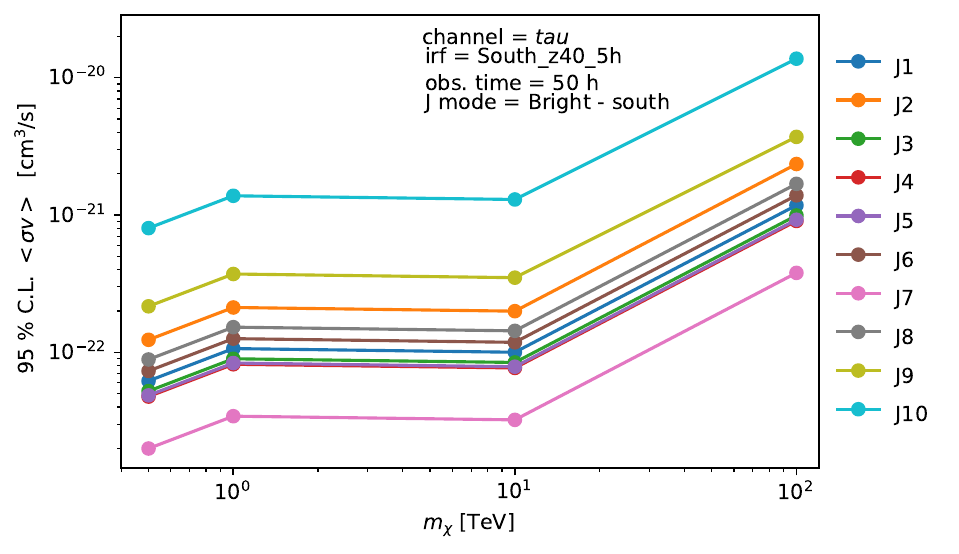}
    \caption{Bright --- South --- $\tau^{+}\tau^{-}$}
    \label{fig:n:t}
  \end{subfigure}
 \caption{CTA 95\% C.L. upper limit on the velocity-weighted cross-section $\braket{\sigma v}$ for different source strengths, array site and annihilation channel.}
 \label{fig:95cl}
\end{figure}

In order to estimate the future sensitivity of CTA in case of no detection, we simply need to search for the value of $\braket{\sigma v}$ for which TS = 2.71. This value of TS corresponds to the chance probability (p-value) associated with 95\% C. L. The calculation is repeated for $m_\chi = 0.5, 1.0, 10.0$, and 100 TeV. Figure~\ref{fig:95cl} shows the lower limit of the sensitivity for each case. The maximum difference between the two sites is 30\% for SM8.


\section{Conclusions}
\label{sec:conclusion}

We analysed the sensitivity of CTA to a signal of dark matter originated from Milky Way satellites. Although in general subhalos provide a faint dark matter signal, their low astrophysical background places them as promising targets for indirect searches.

Using studies that predict the total satellite population of the Milky Way to model the galactic dark matter substructure~\cite{newton, hargis, koposov}, we resolve dark matter subhalos with the \texttt{CLUMPY} code. Because there is little correlation between the radial, mass and concentration parameter distributions derived from the N-body numerical simulations used here, we performed simulations of 500 sky-maps for 10 different substructure modellings (SM$_{i}$, $i \in \{1, …, 10\}$) with the following varying parameters: a) subhalos inner profile, b) spatial distribution of subhalos, c) mass distribution of subhalos, d) total number of subhalos and e) concentration parameter.

For each model, we assess the properties of the average population of the dark matter-brightest subhalos. We also present the distribution of the J-factors from sources. We show that the change in the mass-concentration parameter relation according to reference~\cite{pieri} gives the highest median of the source J-factor. On the other hand, the substructure modelling according to reference~\cite{hargis}, and assuming the lower limit for the number of subhalos expected by the same study, produced the lowest median of the source J-factor.

We proceed with the analysis by calculating the upper (u. l.) and lower limits (l. l.) of the 95 $\%$ C. L. J-factor (J$_{f}$) interval estimated over the 500 brightest subhalos from each model. Assuming sources with such $J_{f}$, we calculate the gamma-ray flux originated from dark matter annihilation into the $\tau^{+}\tau^{-}$ and $b\bar{b}$ channels. This flux was used to estimate the source sensitivity curve, i.e., the parameter space given by the velocity averaged cross-section by the dark matter mass, $\braket{\sigma v}$ $\times$ $m_{\chi}$.

In this step, we assume two major cases: when there is a detection of a dark matter signal with the CTA and when there is not. In the first, we can calculate the true value of $\braket{\sigma v}$ for annihilating dark matter that would give a 5$\sigma$ signal detection. Whereas in the second case, we derived the future $\braket{\sigma v}$ sensitivity at 95\% C.L.

For each dark matter substructure modelling we determine the sensitivity curve for annihilation into the $\tau^{+}\tau^{-}$ and $b\bar{b}$ channels, for sources placed at the north (RA = 180$^{\circ}$, DEC = 45$^{\circ}$) (J2000) and south (RA = 180$^{\circ}$, DEC = -45$^{\circ}$) (J2000) hemispheres. We assume the subhalos have J-factors given by the lower (Faint) and upper (Bright) limits of  J$_{f}$ interval calculated over the 500 brightest subhalos in each model. 

As expected, the model with the highest median J-factor also reached the lowest values of $\braket{\sigma v}$. We show that the $\tau^{+}\tau^{-}$ channel essentially gave the most constrained sensitivity curves and that the parameter space for the north and south hemisphere are roughly equivalent.

The calculated detected or lower limit sensitivity depends significantly on the subhalo model in question. The interval defined between Faint and Bright sources is several orders of magnitudes in some cases. Future studies should consider these results as systematic effects, which will in turn enlarge the uncertainty on the constraints that are imposed on the dark matter model parameters, in particular on the annihilation cross-section. In parallel, an effort should be done to reduce the discrepancy between the substructure models such as to allow a more precise expectation of the possibility of dark matter detection in subhalos when CTA is operational. 



\acknowledgments

Authors acknowledge FAPESP Project 2021/01089-1. MCK acknowledge FAPESP Project 2018/25793-7. AV acknowledge grant 2019/14893-3, São Paulo Research Foundation (FAPESP). Authors acknowledge the National Laboratory for Scientific Computing (LNCC/MCTI,  Brazil) for providing HPC resources of the SDumont supercomputer (http://sdumont.lncc.br). VdS and AV acknowledge CNPq.

\vspace{1cm }
\hspace{-0.8cm}\textbf{Statement of Provenance}
\newline

This is the Accepted Manuscript version of an article accepted for publication in Journal of Cosmology and Astroparticle Physics. Neither SISSA Medialab SrI nor IOP Publishing Ltd is responsible for any errors or omissions in this version of the manuscript or any version derived from it. The Version of Record is available online at https://doi.org/10.1088/1475-7516/2023/03/025.

\newpage

\newpage

\bibliographystyle{JHEP}
\bibliography{references}

\providecommand{\href}[2]{#2}\begingroup\raggedright\begin{thebibliography}{10}

\bibitem{cta2017}
{THE CTA CONSORTIUM, }{\emph{Science with the {C}herenkov {T}elescope {A}rray}
  (2017) } [\href{https://arxiv.org/abs/1709.07997}{{\ttfamily 1709.07997}}].

\bibitem{sdss}
J.E.~Gunn and D.H.~Weinberg, \emph{The sloan digital sky survey},
  \href{https://arxiv.org/abs/astro-ph/9412080}{{\ttfamily astro-ph/9412080}}.

\bibitem{des1}
S.~Koposov et~al., \emph{Beasts of the southern wild: discovery of nine ultra
  faint satellites in the vicinity of the magellanic clouds.},
  {\emph{Astrophysical Journal} {\bfseries 805} (2015) 130}
  [\href{https://arxiv.org/abs/1503.02079}{{\ttfamily 1503.02079}}].

\bibitem{des2}
{THE DES COLLABORATION}, \emph{Eight new {M}ilky {W}ay companions discovered in
  first-year {D}ark {E}nergy {S}urvey data},
  \href{https://doi.org/10.1088/0004-637X/807/1/50}{\emph{Astrophysical
  Journal} {\bfseries 807} (2015) 50}
  [\href{https://arxiv.org/abs/1503.02584}{{\ttfamily 1503.02584}}].

\bibitem{Aquarius}
V.~Springel et~al., \emph{The {A}quarius project: the subhaloes of galactic
  haloes},
  \href{https://doi.org/10.1111/j.1365-2966.2008.14066.x}{\emph{Monthly Notices
  of the Royal Astronomical Society} {\bfseries 391} (2008) 1685}.

\bibitem{elvis}
S.~Garrison-Kimmel, M.~Boylan-Kolchin, J.~Bullock and K.~Lee, \emph{{ELVIS}:
  exploring the local colume in simulations},
  \href{https://doi.org/10.1093/mnras/stt2377}{\emph{Monthly Notices of the
  Royal Astronomy Society} {\bfseries 438} (2014) 2578}.

\bibitem{Zavala:2019gpq}
J.~Zavala and C.S.~Frenk, \emph{{Dark matter haloes and subhaloes}},
  \href{https://doi.org/10.3390/galaxies7040081}{\emph{Galaxies} {\bfseries 7}
  (2019) 81} [\href{https://arxiv.org/abs/1907.11775}{{\ttfamily 1907.11775}}].

\bibitem{Khlopov}
K.~Belotsky, A.~Kirillov and M.~Khlopov, \emph{Gamma-ray evidence for dark
  matter clumps},
  \href{https://doi.org/10.1134/S0202289314010022}{\emph{Gravitation and
  Cosmology} {\bfseries 20} (2012) }.

\bibitem{Coronado-Blazquez:2021avx}
J.~Coronado-Bl\'azquez, M.~Doro, M.A.~S\'anchez-Conde and A.~Aguirre-Santaella,
  \emph{{Sensitivity of the Cherenkov Telescope Array to dark subhalos}},
  \href{https://doi.org/10.1016/j.dark.2021.100845}{\emph{Phys. Dark Univ.}
  {\bfseries 32} (2021) 100845}
  [\href{https://arxiv.org/abs/2101.10003}{{\ttfamily 2101.10003}}].

\bibitem{dmhutten}
M.~H\"utten, C.~Combet, G.~Maier and D.~Maurin, \emph{{Dark matter substructure
  modelling and sensitivity of the Cherenkov Telescope Array to Galactic dark
  halos}}, \href{https://doi.org/10.1088/1475-7516/2016/09/047}{\emph{JCAP}
  {\bfseries 09} (2016) 047}
  [\href{https://arxiv.org/abs/1606.04898}{{\ttfamily 1606.04898}}].

\bibitem{vialacteaII}
M.~Kuhlen, P.~Madau and J.~Silk, \emph{Exploring dark matter with {Milky Way}
  substructure}, {\emph{Science} {\bfseries 325} (2009) 970}.

\bibitem{2011ApJ...736...59Z}
I.~{Zehavi}, Z.~{Zheng}, D.H.~{Weinberg}, M.R.~{Blanton}, N.A.~{Bahcall},
  A.A.~{Berlind} et~al., \emph{{Galaxy Clustering in the Completed SDSS
  Redshift Survey: The Dependence on Color and Luminosity}},
  \href{https://doi.org/10.1088/0004-637X/736/1/59}{\emph{The Astrophysical
  Journal} {\bfseries 736} (2011) 59}
  [\href{https://arxiv.org/abs/1005.2413}{{\ttfamily 1005.2413}}].

\bibitem{2015ApJ...807..152S}
R.A.~{Skibba}, A.L.~{Coil}, A.J.~{Mendez}, M.R.~{Blanton}, A.D.~{Bray},
  R.J.~{Cool} et~al., \emph{{Dark Matter Halo Models of Stellar Mass-dependent
  Galaxy Clustering in PRIMUS+DEEP2 at 0.2>z>1.2}},
  \href{https://doi.org/10.1088/0004-637X/807/2/152}{\emph{The Astrophysical
  Journal} {\bfseries 807} (2015) 152}
  [\href{https://arxiv.org/abs/1503.00731}{{\ttfamily 1503.00731}}].

\bibitem{newton}
O.~{Newton}, M.~{Cautun}, A.~{Jenkins}, C.S.~{Frenk} and J.C.~{Helly},
  \emph{{The total satellite population of the Milky Way}},
  \href{https://doi.org/10.1093/mnras/sty1085}{\emph{Monthly Notices of the
  Royal Astronomy Society} {\bfseries 479} (2018) 2853}
  [\href{https://arxiv.org/abs/1708.04247}{{\ttfamily 1708.04247}}].

\bibitem{hargis}
S.Y.~Kim, A.H.G.~Peter and J.R.~Hargis, \emph{Missing satellites problem:
  completeness corrections to the number of satellite galaxies in the {M}ilky
  {W}ay are consistent with cold dark matter predictions},
  \href{https://doi.org/10.1103/PhysRevLett.121.211302}{\emph{Physical Review
  Letters} {\bfseries 121} (2018) 211302}.

\bibitem{koposov}
S.~Koposov et~al., \emph{The luminosity function of the {M}ilky {W}ay
  satellites}, \href{https://doi.org/10.1086/589911}{\emph{Astrophysical
  Journal} {\bfseries 686} (2008) 279}.

\bibitem{nfw}
J.~Navarro, C.~Frenk and S.~White, \emph{The structure of cold dark matter
  halos}, \href{https://doi.org/10.1086/177173}{\emph{Astrophysical Journal}
  {\bfseries 462} (1995) 563}.

\bibitem{earth}
J.I.~Read, \emph{The local dark matter density},
  \href{https://doi.org/10.1088/0954-3899/41/6/063101}{\emph{Journal of Physics
  G Nuclear Physics} {\bfseries 41} (2014) 063101}
  [\href{https://arxiv.org/abs/1404.1938}{{\ttfamily 1404.1938}}].

\bibitem{diemer}
B.~Diemer and A.V.~Kravtsov, \emph{{A Universal Model for Halo
  Concentrations}},
  \href{https://doi.org/10.1088/0004-637X/799/1/108}{\emph{The Astrophysical
  Journal} {\bfseries 799} (2015) 108}
  [\href{https://arxiv.org/abs/1407.4730}{{\ttfamily 1407.4730}}].

\bibitem{bullock}
J.S.~Bullock et~al., \emph{Profiles of dark haloes: evolution, scatter and
  environment},
  \href{https://doi.org/10.1046/j.1365-8711.2001.04068.x}{\emph{Monthly Notices
  of the Royal Astronomy Society} {\bfseries 321} (2001) 559}
  [\href{https://arxiv.org/abs/astro-ph/9908159}{{\ttfamily
  astro-ph/9908159}}].

\bibitem{sanchez}
M.A.~Sánchez-Conde and F.~Prada, \emph{{The flattening of the
  concentration–mass relation towards low halo masses and its implications
  for the annihilation signal boost}},
  \href{https://doi.org/10.1093/mnras/stu1014}{\emph{Monthly Notices of the
  Royal Astronomical Society} {\bfseries 442} (2014) 2271}
  [\href{https://arxiv.org/abs/https://academic.oup.com/mnras/article-pdf/442/3/2271/3498708/stu1014.pdf}{{\ttfamily
  https://academic.oup.com/mnras/article-pdf/442/3/2271/3498708/stu1014.pdf}}].

\bibitem{moline}
{\'A}.~Molin{\'e}, M.A.~S{\'a}nchez-Conde, S.~Palomares-Ruiz and F.~Prada,
  \emph{Characterization of subhalo structural properties and implications for
  dark matter annihilation signals},
  \href{https://doi.org/10.1093/mnras/stx026}{\emph{Monthly Notices of the
  Royal Astronomical Society} {\bfseries 466} (2017) 4974}
  [\href{https://arxiv.org/abs/1603.04057}{{\ttfamily 1603.04057}}].

\bibitem{pieri}
L.~Pieri, J.~Lavalle, G.~Bertone and E.~Branchini, \emph{Implications of
  high-resolution simulations on indirect dark matter searches},
  \href{https://doi.org/10.1103/PhysRevD.83.023518}{\emph{Physical Review D}
  {\bfseries 83} (2011) 023518}
  [\href{https://arxiv.org/abs/0908.0195}{{\ttfamily 0908.0195}}].

\bibitem{einasto}
A.D.~Ludlow and R.E.~Angulo, \emph{Einasto profiles and the dark matter power
  spectrum}, \href{https://doi.org/10.1093/mnrasl/slw216}{\emph{Monthly Notices
  of the Royal Astronomy Society} {\bfseries 465} (2017) 84}.

\bibitem{Stref:2016uzb}
M.~Stref and J.~Lavalle, \emph{{Modeling dark matter subhalos in a constrained
  galaxy: Global mass and boosted annihilation profiles}},
  \href{https://doi.org/10.1103/PhysRevD.95.063003}{\emph{Phys. Rev. D}
  {\bfseries 95} (2017) 063003}
  [\href{https://arxiv.org/abs/1610.02233}{{\ttfamily 1610.02233}}].

\bibitem{vialactea}
J.~Diemand, M.~Kuhlen and P.~Madau, \emph{Dark matter substructure and
  gamma-ray annihilation in the milky way halo},
  \href{https://doi.org/10.1086/510736}{\emph{The Astrophysical Journal}
  {\bfseries 657} (2007) 262}.

\bibitem{clumpy3}
M.~H{\"u}tten, C.~Combet and D.~Maurin, \emph{{CLUMPY v3}:
  {\ensuremath{\gamma}}-ray and {\ensuremath{\nu}} signals from dark matter at
  all scales}, \href{https://doi.org/10.1016/j.cpc.2018.10.001}{\emph{Computer
  Physics Communications} {\bfseries 235} (2019) 336}
  [\href{https://arxiv.org/abs/1806.08639}{{\ttfamily 1806.08639}}].

\bibitem{ctoolsweb}
J.~Kn\"odlseder et~al., ``The {CTOOL}s website.''
  http://cta.irap.omp.eu/ctools/index.html.

\bibitem{ctools}
J.~Kn{\"o}dlseder et~al., \emph{Gamma{L}ib and ctools. a software framework for
  the analysis of astronomical gamma-ray data},
  \href{https://doi.org/10.1051/0004-6361/201628822}{\emph{Astronomy $\&$
  Astrophysics} {\bfseries 593} (2016) A1}.

\bibitem{2019APh...111...35A}
{THE CTA CONSORTIUM}, \emph{Monte carlo studies for the optimisation of the
  {Cherenkov Telescope Array} layout},
  \href{https://doi.org/10.1016/j.astropartphys.2019.04.001}{\emph{Astroparticle
  Physics} {\bfseries 111} (2019) 35}
  [\href{https://arxiv.org/abs/1904.01426}{{\ttfamily 1904.01426}}].

\end{thebibliography}\endgroup


\newpage


\end{document}